\documentclass[aps,prl,reprint,letterpaper,amsmath,amssymb, superscriptaddress]{revtex4-1}
\usepackage{graphicx}
\usepackage[absolute]{textpos}

\begin{document}
\begin{textblock*}{8.5in}(0.1in,0.25in)
\begin{center}
PHYSICAL REVIEW B \textbf{94}, 241202(R) (2016)
\end{center}
\end{textblock*}

\title{Extended pump-probe Faraday rotation spectroscopy of the submicrosecond electron spin dynamics in $n$-type GaAs}

\author{V. V. Belykh}
\email[]{vasilii.belykh@tu-dortmund.de}
\affiliation{Experimentelle Physik 2, Technische Universit\"{a}t Dortmund, D-44221 Dortmund, Germany}
\affiliation{P.N. Lebedev Physical Institute of the Russian Academy of Sciences, 119991 Moscow, Russia}
\author{E. Evers}
\affiliation{Experimentelle Physik 2, Technische Universit\"{a}t Dortmund, D-44221 Dortmund, Germany}
\author{D. R. Yakovlev}
\affiliation{Experimentelle Physik 2, Technische Universit\"{a}t Dortmund, D-44221 Dortmund, Germany}
\affiliation{Ioffe Institute, Russian Academy of Sciences, 194021 St. Petersburg, Russia}
\author{F. Fobbe}
\affiliation{Experimentelle Physik 2, Technische Universit\"{a}t Dortmund, D-44221 Dortmund, Germany}
\author{A. Greilich}
\affiliation{Experimentelle Physik 2, Technische Universit\"{a}t Dortmund, D-44221 Dortmund, Germany}
\author{M. Bayer}
\affiliation{Experimentelle Physik 2, Technische Universit\"{a}t Dortmund, D-44221 Dortmund, Germany}
\affiliation{Ioffe Institute, Russian Academy of Sciences, 194021 St. Petersburg, Russia}

\received{28 October 2016} \published{14 December 2016}

\begin{abstract}
We develop an extended pump-probe Faraday rotation technique to study submicrosecond electron spin dynamics with picosecond time resolution in a wide range of magnetic fields. The electron spin dephasing time $T_2^*$ and the longitudinal spin relaxation time $T_1$, both approaching $250$~ns in weak fields, are measured thereby in $n$-type bulk GaAs. By tailoring the pump pulse train through increasing the contained number of pulses, the buildup of resonant spin amplification is demonstrated for the electron spin polarization. The spin precession amplitude in high magnetic fields applied in the Voigt geometry shows a non-monotonic dynamics deviating strongly from a mono-exponential decay and revealing slow beatings. The beatings indicate a two spin component behavior with a $g$-factor difference of $\Delta g \sim 4\times10^{-4}$, much smaller than the $\Delta g$ expected for free and donor-bound electrons. This $g$-factor variation indicates efficient, but incomplete spin exchange averaging.
\\
\\
\doi{10.1103/PhysRevB.94.241202}
\end{abstract}

\maketitle

Initialized electron spins in semiconductors undergo a complex dynamics depending on external magnetic field, interaction with other charge carriers and nuclei, spin-orbit interaction, etc. Knowledge of the resulting spin dynamics provides information on these interactions and related spin properties such as $g$ factors and relaxation times which are important for basic research and application in information technologies. Commonly, information on spin properties is mostly obtained from resonance techniques like electron paramagnetic resonance, optically-detected magnetic resonance, spin-flip Raman scattering, or polarized photoluminescence (Hanle effect). The development of pump-probe Faraday/Kerr rotation spectroscopy has facilitated exploration of the coherent spin dynamics, in particular the Larmor spin precession around a magnetic field, with picosecond temporal resolution and opened new ways for spin control and manipulation \cite{Awschalom2002, Dyakonov2008,Slavcheva2010,Greilich2006,Greilich2007}.

The main limitation imposed on the standard pump-probe technique is the restricted time range that can be monitored. This restriction comes from  the finite length of mechanical delay lines for the pump-probe delay limiting this time range to a few nanoseconds, which can be too short to address the carrier spin dynamics in semiconductors. To evaluate longer spin dephasing times the resonant spin amplification (RSA) technique \cite{Kikkawa1998,Yugova2012} can be used, which, however, does not provide detailed insight into complex spin dynamics such as a nonexponential decay of spin polarization. Also, the longitudinal spin relaxation characterized by the $T_1$ time typically exceeds the nanosecond range, so that indirect optical techniques like the spin inertia method \cite{Heisterkamp2015} have to be used, again with limited access to  nontrivial spin dynamics.

Here we extend the standard pump-probe Faraday rotation (PPFR) technique to address a much longer time range by employing a tailored pump pulse sequence, while maintaining picosecond time resolution. The technique is applied to the submicrosecond electron spin dynamics in bulk $n$-type GaAs. The spin dephasing time $T_2^*$ measured thereby from the dynamics of spin precession at low magnetic fields is in agreement with data recorded from the Hanle effect \cite{Dzhioev2002a,Furis2007}, RSA \cite{Kikkawa1998} and spin noise spectroscopy \cite{Crooker2009, Romer2010}. However, at increased magnetic fields the spin precession decay becomes nonexponential, a behaviour hardly accessible in detail by other methods. This peculiar dynamics is characterized by slow beatings between two electron subensembles shedding light on the spin exchange averaging in semiconductors \cite{Paget1981}. Further, we demonstrate the buildup of electron spin polarization in the RSA regime with increasing number of pump pulses in a train, which provides an alternative approach for measuring $T_2^*$. In longitudinal magnetic fields (Faraday geometry) the longitudinal spin relaxation time $T_1$ is measured in a wide range of fields.


The sample under study is a 350-$\mu$m thick GaAs epitaxial layer doped with Si to provide a donor concentration $n_{\rm D} = 1.4\times10^{16}$~cm$^{-3}$. Some results are also presented for samples with $n_{\rm D}=3.7$ and $7.1\times10^{16}$~cm$^{-3}$ having thicknesses of 170~$\mu$m. The samples were placed in a split-coil magnetocryostat in contact with helium gas at a temperature $T=6$~K. Magnetic fields up to 6~T were applied either parallel (Faraday geometry) or perpendicular (Voigt geometry) to the light propagation vector (and sample growth axis).

\begin{figure}
\includegraphics[width=1\columnwidth]{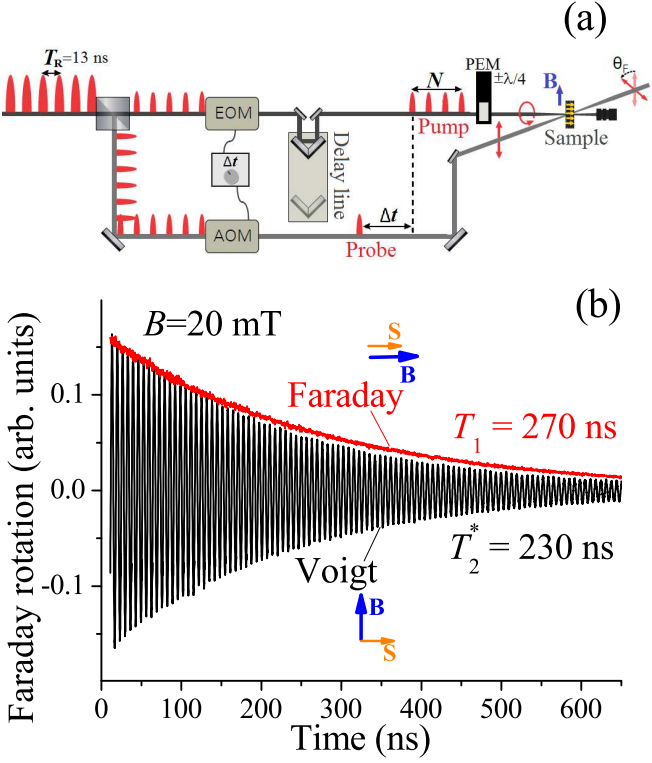}
\caption{(a) Scheme of extended PPFR experiment. (b) Dynamics of Faraday rotation signal for $B = 20$~mT applied in Voigt (black line) and Faraday (red line) geometry. $n_{\rm D}=1.4\times10^{16}$~cm$^{-3}$.
}
\label{FigKin}
\end{figure}

The extended PPFR technique [Fig.~\ref{FigKin}(a)] is a modification of the common pump-probe Faraday rotation technique, where circularly-polarized pump pulses generate carrier spin polarization, which is then probed by the Faraday rotation of linearly-polarized probe pulses after transmission through the sample. The temporal evolution of the spin polarization is traced by varying the time delay between pump and probe pulses. In order to go for long time delays and to have flexibility with setting excitation protocols we implement pulse picking for both
pump and probe laser beams.

We use a Ti:Sapphire laser emitting a train of 2~ps pulses with a repetition rate of 76~MHz (repetition period $T_\text{R}=13.1$~ns). The laser output is split into pump and probe [Fig.~\ref{FigKin}(a)].
In the pump path an electro-optical modulator (EOM) is installed to select trains of $N$ pulses (from 1 to about 100) separated by $T_\text{R}$ with arbitrarily long delay between the trains.
An acousto-optical light modulator (AOM) in the probe path is used to  select single pulses at the required delay after the pump train. Electronic variation of the delay between the synchronized AOM and EOM (also synchronized with the laser) allows for a coarse change of the delay between the pump pulse sequence and the probe pulse in steps of $T_\text{R}$, providing the desired time range. In addition, a mechanical delay line in the pump path allows for fine delay variation up to $T_\text{R}$. In this way the Faraday rotation dynamics could be measured over a microsecond time range with still  2~ps time resolution. Except for RSA experiments with variable $N$, we use here trains of $N = 8$ successive pump pulses, applied at a train repetition period of $80 T_\text{R}=1.05$~$\mu$s. To perform  synchronous detection and to avoid nuclear polarization, the polarization of the pump was modulated between $\sigma^{+}$ and $\sigma^{-}$ by a photo-elastic modulator (PEM) operated at a frequency of 84 kHz.
The laser wavelength was set to 825~nm (827 and 829 nm for the samples with $n_{\rm D}=3.7$ and $7.1\times10^{16}$~cm$^{-3}$, respectively), below the GaAs band gap, to avoid complete absorption of the probe. The average pump power for the protocol ``8 out of 80 pulses'' was $P=0.1$~mW. The diameter of the pump spot on the sample was about 100~$\mu$m.

The black line in Fig.~\ref{FigKin}(b) shows the Faraday rotation dynamics measured with the extended PPFR technique in a magnetic field $B_{\perp}=20$~mT applied in the Voigt geometry for the sample with $n_{\rm D}=1.4\times10^{16}$~cm$^{-3}$. The signal shows oscillatory behaviour caused by the electron spin precession at frequency $\omega=|g|\mu_\text{B}B_{\perp}/\hbar$, where $g=-0.44$ is the electron $g$ factor in bulk GaAs \cite{Madelung1996} and $\mu_\text{B}$ is the Bohr magneton. The oscillation amplitude decays exponentially with the ensemble spin dephasing time $T_2^*=230$~ns. Note that the separation between the pump pulse trains is $80 T_\text{R} \approx 1.05$~$\mu$s~$\gg T_2^*$. The measured $T_2^*$ is close to the values obtained from  RSA \cite{Kikkawa1998}, Hanle \cite{Dzhioev2002a,Furis2007}, and spin noise \cite{Crooker2009, Romer2010} experiments at $B\approx 0$.

\begin{figure}
\includegraphics[width=1\columnwidth]{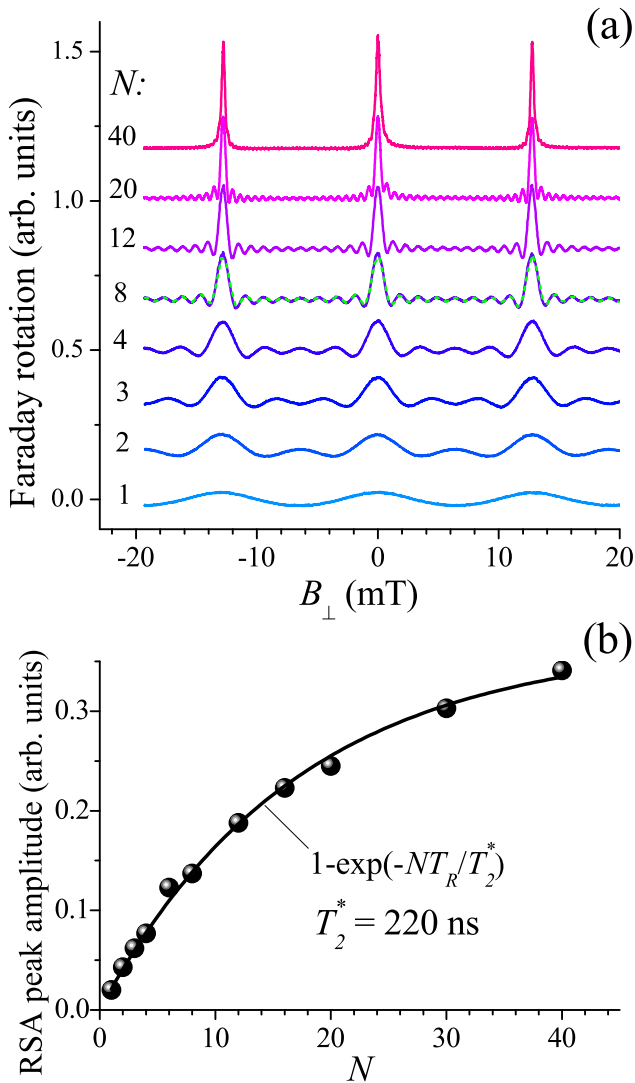}
\caption{(a) Resonant spin amplification curves for different numbers of pump pulses $N$ before the probe pulse that is delayed by 12.9~ns relative to the last pump in a train. The green dashed line shows a fit to the experiment with Eq.~\eqref{EqRSAN} for $N=8$. (b) Dependence of the RSA peak amplitude at $B_{\perp}=0$ on the number of pump pulses. The solid line gives a fit to the experiment. $n_{\rm D}=1.4\times10^{16}$~cm$^{-3}$.}
\label{FigRSA}
\end{figure}

The extended PPFR technique allows one to study the buildup of electron spin polarization in the RSA regime with increasing number of pump pulses $N$ in the train that precedes the probe pulse. The delay between the probe and the last pump pulse in a train is set to $\Delta t=T_\text{R}-0.2$~ns $\approx 12.9$~ns [Fig. \ref{FigKin}(a)]. The Faraday rotation signal is measured as a function of magnetic field applied in the Voigt geometry. Figure~\ref{FigRSA}(a) shows RSA curves for different $N$. For $N=1$, the RSA curve shows a sinusoidal oscillation with period $\Delta B_{\perp} = 2\pi\hbar/(\Delta t |g|\mu_\text{B})$. With increasing $N$, resonances at magnetic fields $B_q=2\pi\hbar q/(T_\text{R} |g|\mu_\text{B})\approx q \times 13$~mT appear, where $q$ is an integer. These resonances correspond to $q$ complete spin revolutions between subsequent pump pulses. With increasing $N$ the main RSA resonances increase in amplitude and narrow resulting in a curve with sharp peaks.
Between the RSA resonances $N-1$ oscillations are seen corresponding to interference of spin precessions initiated by different pump pulses within a train. With increasing $N$ the oscillations become faint and disappear for $N \to \infty$, so that they are not seen in standard RSA curves.

The RSA curve can be described by superposition of $N$ damped oscillations \cite{Yugova2012}:
\begin{equation}
S = S_0 \Sigma_{q=0}^{N-1} \cos[\omega(\Delta t + q  T_\text{R})]\exp\left(-\frac{\Delta t + q T_\text{R}}{T_2^*}\right),
\label{EqRSAN}
\end{equation}
where $S_0$ is the spin polarization induced by a single pump pulse and the magnetic field dependence is represented by $\omega=|g|\mu_\text{B}B_{\perp}/\hbar$.
The experimental dependencies are perfectly reproduced with Eq.~\eqref{EqRSAN}, the corresponding fit is shown in Fig.~\ref{FigRSA}(a) by the green dashed line for $N=8$.

The half width at half maximum (HWHM) of the RSA peak for $N\gg T_2^*/T_\text{R}$ saturates at $\delta B_{\perp} = \hbar/(|g|\mu_\text{B} T_2^*)$, which gives the established way to evaluate $T_2^*$ \cite{Kikkawa1998}.  In our case the HWHM for $N\rightarrow\infty$ is 0.15~mT corresponding to $T_2^*\approx 170$~ns, underestimating somewhat the value from a direct measurement [see Fig.~\ref{FigKin}(b)], but being still in reasonable agreement.

The dependence of the RSA peak amplitude on $N$ gives another way to determine $T_2^*$. Indeed, according to Eq.~\eqref{EqRSAN} $S(\omega=0)\propto 1-\exp(-NT_\text{R}/T_2^*)$, which well describes the experimental dependence in Fig.~\ref{FigRSA}(b). The fit gives $T_2^*\approx220$~ns, in good agreement with the extended PPFR measurement from Fig.~\ref{FigKin}(b).

\begin{figure*}
\includegraphics[width=2\columnwidth]{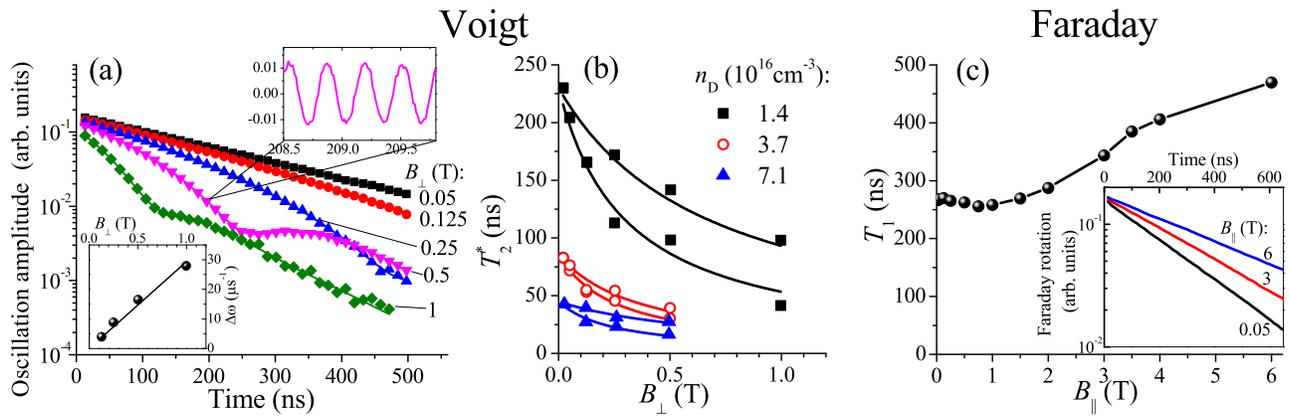}
\caption{(a) Dynamics of Faraday rotation oscillation amplitude for different magnetic fields applied in Voigt geometry. The upper inset shows a closeup of the spin precession. The lower inset shows magnetic field dependence of the beating frequency $\Delta\omega$; the line is a linear fit. (b) Magnetic field dependence of the transversal spin dephasing times $T_2^*$ of the two beating components for three samples with different doping concentrations. (c) Dependence of the longitudinal spin relaxation time $T_1$ on the magnetic field applied in Faraday geometry. The inset shows corresponding dynamics of the Faraday rotation. The lines in panels (a) and (b) show fits to the experimental data (see text), lines in panel (c) are guides to the eye. The data in panels (a) and (c) correspond to the sample with $n_{\rm D}=1.4\times10^{16}$~cm$^{-3}$.}
\label{FigHighB}
\end{figure*}

Thus, the RSA technique gives the correct value of $T_2^*$ for simple exponential dephasing of the spin polarization. At increased magnetic field, basic characteristics of the long-lasting spin dynamics can also be extracted from the RSA technique \cite{Kikkawa1998} as well as from the recently developed heterodyne detection of spin noise \cite{Cronenberger2016}. However, as we will show below, at increased $B_{\perp}$ the dynamics reveals peculiarities hardly accessible by indirect methods. We turn now to measurements of the electron spin dynamics at long delays for pump trains containing 8 pulses. The dynamics are measured for different magnetic fields applied in the Voigt geometry. The precise magnetic field strength was adjusted in a range of few mT around the given value to meet the RSA condition for maximal signal. Figure~\ref{FigHighB}(a) shows the time dependence of the Faraday rotation oscillation amplitude in steps of $13.1$~ns, i.e. of the envelope of the spin precession dynamics [upper inset in Fig.~\ref{FigHighB}(a)], for the sample with $n_{\rm D}=1.4\times10^{16}$~cm$^{-3}$.

At $B_{\perp}=50$~mT the spin precession amplitude shows an exponential decay [Fig.~\ref{FigHighB}(a)]. At $B_{\perp}=125$~mT the dynamics accelerates at longer times, thereby significantly deviating from an exponential decay. Further field increase reveals a dip in the dynamics, which shifts to shorter times with increasing field, while the overall dynamics accelerate.

The dip in the dynamics originates from beatings of the signals from spin subensembles with a small difference in precession frequency $\Delta\omega$. Indeed, the amplitude dynamics are well fitted with a precession amplitude (envelope) of the sum of two oscillating components $A_1\exp(-t/\tau_1)\cos(\omega t) + A_2\exp(-t/\tau_2)\cos[(\omega + \Delta \omega)t]$ as shown by the lines in Fig.~\ref{FigHighB}(a).
From the fits we determine the $\Delta\omega$, which scales linearly with magnetic field [see lower inset in Fig.~\ref{FigHighB}(a)]. This suggests that $\Delta\omega$ arises from different $g$ factors of two electron subensembles, so that $\Delta\omega=\Delta g\mu_\text{B}B/\hbar$ with $\Delta g \approx 4\times 10^{-4}$.

The magnetic field dependencies of the decay times $T_2^*$ for both subensembles are shown in Fig.~\ref{FigHighB}(b) by the black squares. Their decrease with increasing field is related to the $g$-factor spread $\delta g$ within each subensemble, described by the equation $1/T_2^*(B_{\perp})=1/T_2^*(0)+\delta g\mu_\text{B}B_{\perp}/\hbar$ \cite{Belykh2015}. Corresponding fits are shown by the solid lines in Fig.~\ref{FigHighB}(b) and give $\delta g$ of $2\times10^{-4}$ and $1\times10^{-4}$ for the two subensembles in the sample with $n_{\rm D}=1.4\times10^{16}$~cm$^{-3}$.

The spin precession dynamics for the samples with higher doping concentrations of $3.7 \times 10^{16}$~cm$^{-3}$ and $7.1 \times 10^{16}$~cm$^{-3}$ give spin dephasing times of $80$ and $40$~ns, respectively, at low $B_{\perp}$, while at higher magnetic fields they also reveal slow beatings, corresponding to $\Delta g=1.9 \times10^{-3}$ and $1.6 \times10^{-3}$, respectively. The magnetic field dependencies of the decay times of the two components for these samples are included in Fig.~\ref{FigHighB}(b).

The electron concentrations in the studied samples are close to the metal-insulator transition ($\sim 2\times10^{16}$~cm$^{-3}$) \cite{Dzhioev2002a}. Therefore, it seems reasonable to attribute the two electron subensembles to free and donor-bound electrons. One can estimate the $g$ factor difference $\Delta g_0$ for free and bound electrons from the difference in their transition energies of $\sim 6$~meV \cite{Milnes1973} using the Roth-Lax-Zwerdling equation \cite{Roth1959}. The result is $\Delta g_0 \sim 10^{-2}$, considerably larger than the measured $\Delta g \approx 4 \times 10^{-4}$. On the other hand, the exchange interaction between free and bound electrons provides an efficient averaging mechanism \cite{Paget1981}. This mechanism can be qualitatively understood as frequent spin exchange between free and bound electrons by scattering, and it is analogous to the motional narrowing described in Ref.~\cite{Pines1955}. As a result, spin precession occurs on a \emph{single} average frequency.
However, spatial inhomogeneity in the donor distribution may result in a broadening of the frequency distribution. In particular, one may consider different spatial domains of free and bound electron concentrations.

The dispersion of the free electron $g$ factor near the GaAs band gap that is approximately given by $g(E) = -0.44 + \beta E$, where $\beta = 6.3$~eV$^{-1}$ \cite{Yang1993} provides an additional mechanism of $g$ factor broadening. In particular, at $T=6$~K the temperature broadening is expected to be $\delta g_{T}\approx 3 \times 10^{-3}$, which is an order of magnitude larger than the measured $\Delta g \approx 4 \times 10^{-4}$. Obviously, the spin exchange averaging mechanism is responsible for that.

It is straightforward to apply the extended PPFR technique to measuring the longitudinal spin relaxation time $T_1$ in a magnetic field $B_{\parallel}$ applied in Faraday geometry. The red line in Fig.~\ref{FigKin}(b) shows the dynamics of the Faraday rotation at $B_{\parallel}=20$~mT for the sample with $n_{\rm D}=1.4\times10^{16}$~cm$^{-3}$. The signal shows a monoexponential decay without oscillations with $T_1=270$~ns, close to the measured $T_2^*=230$~ns at low magnetic fields. Note that for $B\rightarrow 0$ we expect $T_2^*=T_2=T_1$. The decay is monoexponential in the whole range of magnetic fields $B_{\parallel} \leqslant 6$~T [inset in Fig.~\ref{FigHighB}(c)], and the corresponding decay time $T_1$ increases with $B_{\parallel}$ above $\sim 1.5$~T [Fig.~\ref{FigHighB}(c)]. The suppression of spin relaxation by a longitudinal magnetic field is much weaker than reported for bulk GaAs with lower donor concentrations (well below the metal-insulator transition) \cite{Colton2004,Linpeng2016,Fu2006}. Indeed, for the studied donor concentration close to the metal-insulator transition, the electron spin relaxation is dominated by the Dyakonov-Perel and anisotropic exchange mechanisms \cite{Dzhioev2002a}, which are less suppressed by the external field compared to the spin relaxation due to the electron hyperfine interaction with nuclei that is dominating factor at low doping concentrations \cite{Linpeng2016}. We are not aware of reports on the magnetic field dependence of $T_1$ near the metal-insulator transition and at higher electron densities.

In conclusion, we have developed an extended pump-probe Faraday rotation technique and demonstrated its potential in studying electron spin dynamics with picosecond resolution in a wide temporal range up to 1~$\mu$s and potentially longer. This enables direct measurement of the spin dephasing time $T_2^*$ and longitudinal spin relaxation time $T_1$ for carriers in arbitrary magnetic fields. The technique can be used for high sensitivity spectroscopy of $g$ factors barely accessible by other methods, e. g. RSA. The possibility of varying the pump pulse train composition from single to multiple pulses provides access to the detailed process of electron spin synchronization under periodic laser excitation.

\begin{acknowledgements}
We are grateful to M.~M.~Glazov, V.~L.~Korenev, and A.~V.~Poshakinskiy for valuable advices and discussions and to S.~A.~Crooker for providing the samples and discussions. We acknowledge the financial support of the Deutsche Forschungsgemeinschaft in the frame of the ICRC TRR 160 and the Russian Science Foundation (Grant No. 14-42-00015). Also, the support by the BMBF project Q.com-HL is appreciated.
\end{acknowledgements}

\end{document}